\documentclass[twocolumn, prl]{revtex4}
\usepackage{graphicx}

\begin{document}
%\draft
\title{The time-energy certainty relation}
\author{Zbyszek P. Karkuszewski\footnote{e-mail: zbyszek@pamir.if.uj.edu.pl}}

\affiliation{
Los Alamos National Laboratory, Los Alamos, NM 87545, USA\\
and\\
Institute of Physics, Jagiellonian University, Cracow, Poland\\
}

\date{\today}

\begin{abstract}
All energy measurements of a quantum system are prone to inaccuracies.
In particular, if such measurements are carried over a finite period
of time the accuracy of the result is affected by the length of 
that period. Here I show an upper bound on such inaccuracies and
point out that the bound can be arbitrarily small if many copies
of the system are available.

LAUR-05-9224
\end{abstract}

\maketitle

\section{Introduction}
Since the onset of quantum mechanics the uncertainty principles
have played the central role in attempts to comprehend this
counterintuitive theory. The importance of these principles
springs from the search for limitations of the theory. They tell us
what is possible and what is fundamentally unachievable in the quantum
world.  
Although the mathematical structure of quantum mechanics is well understood, 
its physical interpretation and applicability to the phenomena of the 
observed world is still debated. In particular, there is some confusion
about the status of various time-energy uncertainty relations present
in the textbooks on quantum theory \cite{Schiff,Messiah}. 
In this work I will focus on one important question that arises naturally in the
context of the time-energy uncertainties: How accurate the energy measurement
can be if it is carried over a small period of time?
It has been noticed that there is no fundamental lower bound on such an
accuracy no matter how short the measurement \cite{Aharonov1,Asher}. In the course
of this Letter I derive an upper bound on the accuracy, which the energy of
a quantum system can be determined with after a short measurement. To avoid
further confusion on the subject, the derivation, which is simple enough, is presented in 
reasonable detail within the scope of this work. The final result is given in
Eq. (\ref{ieq2}).

Lets consider a general setup where an evolution of a quantum system is governed by 
an unknown time-independent Hamiltonian of discrete spectrum $\hat H$. 
Any state of the system can be written in terms of the corresponding energy eigenstates $|k\rangle$, 
but here I will consider only the states that are dominated only by a finite number $K$ of the
nondegenerate eigenstates. Thus, the initial state of the system is expressed as 
\begin{equation}
|\psi_0\rangle = \sum_{k=1}^K \alpha_k |k\rangle,
\label{initstate}
\end{equation}
where the amplitudes $\alpha_k$, the eigenstates $|k\rangle$  and $K$ are initially unknown. 
Later it will become clear that some states that are spanned over infinite number of the 
eigenstates can also be included into the consideration. 

I will be concerned only with the states at the discrete time steps $t=n\delta t, n=0,1,...,N\ge K$, 
that are denoted by $|\psi_n\rangle$. In order to determine the eigenenergies $\hbar \omega_k$ 
driving the evolution of $|\psi\rangle$ one can measure the time autocorrelation function
\begin{equation}
c_n \equiv \langle \psi_0|\psi_n\rangle = \sum_{k=1}^K d_k e^{-i\omega_k n\delta t},
\label{autocorrelation}
\end{equation}
with $d_k\equiv |\alpha_k|^2$. This quantity can certainly be measured since
the wave function itself (not just the wave function modulus squared) can be measured with quantum tomography \cite{Raymer}.
The eigenfrequencies are extracted from the eigenvalue problem for the
evolution operator $\hat U$, that evolves the system by one time step $\delta t$,
i.e. $\hat U |\psi_n\rangle = |\psi_{n+1}\rangle$,
\begin{equation}
\hat U |k\rangle = e^{-i \omega_k \delta t}|k\rangle.
\label{eigeq}
\end{equation}
To find a matrix representation of the last equation it is convenient to
arrange the first $N$ states $|\psi_n\rangle$ as columns of $K\times N$ matrix $P$.
Inserting unity $PP^{-1}$ before states $|k\rangle$ in Eq. (\ref{eigeq}),
and multiplying by $P^\dagger$ from the left results in the generalized eigenvalue
problem
\begin{equation}
 U' |k'\rangle = e^{-i \omega_k \delta t} S |k'\rangle,
\label{geigeq}
\end{equation}
where $U'\equiv P^\dagger \hat U P$ and $S\equiv P^\dagger P$ are $N\times N$ matrices 
with matrix elements expressed solely in
terms of the autocorrelation function values, $U'_{mn}=c_{n-m+1}$ and
$S_{mn}=c_{n-m}$. The new eigenvectors are $|k'\rangle \equiv P^{-1}|k\rangle$.

At this point the problem of extracting all the energy related information of the system
in the initial state $|\psi_0\rangle$ is solved. Given the measured values of the
autocorrelation function $c_n$, one finds the number $K$ of eigenstates involved
as a rank of the matrix $S$, eigenfrequencies from the solution of Eq.
(\ref{geigeq}) and coefficients $d_k$ by solving linear set of equations
(\ref{autocorrelation}) with now known $K$ and $\omega_k$.
With known $K$, $d_k$ and $\omega_k$ finding the expectation energy of the system
or the corresponding standard deviation is trivial.
The method presented above goes under the name of harmonic inversion and its variations
has been used by scientists on many occasions \cite{Prony,Neuhauser,Taylor1,Taylor2}.
You can find an introduction to the method in Ref.\cite{Marple}. 

\section{Inaccuracies}
The analysis, so far, did not account for the inevitable measurement inaccuracies.
The experimental values of the autocorrelation function differ from the
unperturbed values that enter Eq.(\ref{autocorrelation}) and this difference
will affect the computed values of the desired eigenfrequencies.
The effect of the experimental imperfectness will be further amplified by the
requirement that the all measuring is done in a very short time interval
$T=N\delta t$, which satisfies $T\ll 2\pi/\Delta\omega_{max}$, with 
$\Delta\omega_{max}\equiv \max_{j,k}\{\omega_j-\omega_k\}$
being the largest possible frequency difference in Eq. (\ref{autocorrelation}).
In other words the total measurement time $T$ is much shorter than the
shortest characteristic time interval of the quantum evolution.

Lets denote the measured value of the autocorrelation function by
$\tilde c_n\equiv c_n + \eta_n$, where $\eta_n$ contains all the statistical inaccuracies
like noise and effect of finite number of the quantum system copies available. The
systematic errors can be, in principle, eliminated and I assume they are
absent. Also a time instant $n\delta t$, at which the measurement of $\tilde c_n$ is performed,
can be estimated with only finite precision, but in this work it assumes a precise value 
$n\delta t$, while the related inaccuracy is relegated to $\eta_n$. 

It is possible that even some states spanned over infinite number of energy
eigenstates can be considered. One would pick $K<\infty$ most important
eigenstates i.e. with the greatest $d_k$'s and treat the remaining part as a
noise contributing to $\eta_n$. However, this contribution would have a
systematic character rather than statistical and should account for only a
negligible part of the total error $\eta_n$.

To summarize the discussion of sources of the inaccuracies observe that all
of them are contained in complex noise terms $\eta_n$. In the absence of systematic
errors, the noise level can be made arbitrarily small by increasing the number of copies
of the quantum system undergoing the measurement. The time-energy uncertainty
relation for a measurement on just one quantum system is introduced in
\cite{Aharonov2}.
The largest magnitude of $\eta_n$ is denoted by $\eta_{max}$.

In order to see the influence of the noise on the extracted eigenfrequencies
one must solve the generalized eigenvalue problem (\ref{geigeq}) with
corrupted by noise matrices $\tilde U'$ and $\tilde S$. To do this lets first
diagonalize the matrix $S$. $S$ is a Hermitian non-negative defined $N\times N$
matrix. It has exactly $K$ positive eigenvalues $\lambda_1\ge\lambda_2\ge
...\ge\lambda_K>\lambda_{K+1}=...=\lambda_N=0$. The smallest positive
eigenvalue will be denoted by $\lambda_{min}\equiv\lambda_K$. The perturbed matrix 
$\tilde S$ is still Hermitian and its eigenvalues differ from the unperturbed ones by at
most $N\eta_{max}$, $|\tilde\lambda_k-\lambda_k|\le N\eta_{max}$. Thus, it is
possible to tell the number of positive eigenvalues of $S$ only for limited
level of noise
\begin{equation}
\eta_{max} < \frac{\lambda_{min}}{2N},
\end{equation}
where the factor $1/2$ comes from the fact that zero eigenvalues are also
perturbed up to $N\eta_{max}$. This is the condition one must fulfill to
extract all $K$ frequencies that enter Eq. (\ref{autocorrelation}).

In the next step one has to truncate the space spanned by eigenvectors of
$\tilde S$ only to those that correspond to $K$ positive eigenvalues, so that
the inverse of the truncated matrix $\tilde S$ could be found. The eigenvectors to
positive eigenvalues of  $\tilde S$ are arranged as a columns of $N\times K$ unitary 
matrix $\tilde Q$.
The diagonal $K\times K$ matrix $\tilde D\equiv \tilde Q^\dagger \tilde S
\tilde Q$ is nonsingular and can be inverted. Also the $\tilde U'$ is truncated to
$\tilde Q^\dagger \tilde U' \tilde Q$. The Eq. (\ref{geigeq}) takes the
form an ordinary eigenvalue problem
\begin{equation}
\tilde D^{-1} \tilde Q^\dagger \tilde U' \tilde Q \tilde Q^\dagger |k'\rangle = 
e^{-i \omega_k \delta t} \tilde Q^\dagger |k'\rangle.
\label{oeigeq}
\end{equation}

In general, for an eigenvalue equation $A X = X \Lambda$, where $X$ is a
matrix of eigenvectors of $A$ and $\Lambda$ is the diagonal matrix of
corresponding eigenvalues, the perturbation $\delta A$ of $A$ results in perturbation 
$\delta \Lambda$ of $\Lambda$ and $X^{-1}(A+\delta A)X = \Lambda + \delta \Lambda$.
It follows that $\delta \Lambda = X^{-1}\delta A X$ and using matrix 2-norm
$\|\delta \Lambda\|\le\|X^{-1}\| \|X\| \|\delta A\|\equiv \kappa(X)\|\delta
A\|$, with $\kappa(X)$ being the condition number of eigenvector matrix $X$
\cite{errorestimate}.

Similarly, to estimate the influence of the perturbations on the eigenvalues 
in Eq. (\ref{oeigeq}) lets rewrite the matrix on the left hand side as follows
\begin{eqnarray}
(D^{-1} + \delta D^{-1}) \tilde Q^\dagger (U' + \delta U')\tilde Q &\approx&
D^{-1}\tilde Q^\dagger U'\tilde Q \nonumber \\
+ \delta D^{-1} \tilde Q^\dagger U' \tilde Q &+& D^{-1} \tilde Q^\dagger \delta U' \tilde Q, 
\label{rosp}
\end{eqnarray}
where $\delta D^{-1}\equiv N\eta_{max} \mbox{diag}(1/\lambda_1^2,...,1/\lambda_K^2)$
and $\delta U'$ has matrix elements $\eta_{n-m+1}$.
The first term on the right hand side of Eq. (\ref{rosp}) has complex eigenvalues on the
unit circle, which means that $i$-th row of $\tilde Q^\dagger U' \tilde Q$ must be
of order of $\lambda_i$ to cancel the factor $1/\lambda_i$ of $D^{-1}$. The
second term must then have matrix elements at most of order of
$N\eta_{max}/\lambda_{min}$. The matrix elements of the third term are clearly
of order of $\eta_{max}/\lambda_{min}$. The last two terms combined give an
estimate of an error made when computing the eigenvalues of Eq. (\ref{oeigeq})
$$
|e^{-i\tilde \omega_k \delta t} - e^{-i\omega_k\delta t}| \le \kappa(P) \frac{K(N+1)\eta_{max}}{\lambda_{min}}
$$
where $\kappa(P)$ is a condition number for matrix $P$ and the fraction is an
estimate of the Frobenius norm, the upper bound for 2-norm, of the two last matrices in 
Eq. (\ref{rosp}).
The condition number for $\tilde Q$ matrix is 1, because $\tilde Q$ is
unitary, and is skipped.  

The last inequality gives an estimate of frequency error made when evolution
time step $\delta t$ is very short i.e. $T\ll 2\pi/\Delta\omega_{max}$. Namely,
\begin{equation}
|\tilde \omega_k - \omega_k|\delta t \le
\kappa(P)\frac{K(N+1)}{\lambda_{min}}\eta_{max}.
\label{ieq1}
\end{equation}
Multiplying both sides above by $N\hbar$ one gets something opposite to a
time-energy uncertainty relation, the upper bound rather than lower bound for
the measured energy inaccuracies. This is why I refer to it as the
time-energy certainty relation. The important remark is that the error of the
measured energies can be kept arbitrarily small, no matter how short the
measurement was, by reducing the noise level $\eta_{max}$. It is assumed that
the condition number $\kappa(P)$ is finite, as will be shown in the following.

The certainty relation as presented in Eq. (\ref{ieq1}) has little use,
except perhaps a philosophical value, since
it still contains rather cryptic quantities $\kappa(P)$ and $\lambda_{min}$.
The prescription for calculating $\kappa(P)$ is provided in the next
paragraph.

Let as take a closer look at the $K\times N$ matrix $P$. As mentioned in the
introduction the matrix columns are made of components of vectors $|\psi_n\rangle$,
$P_{kn}=\alpha_k e^{-i\omega_k n\delta t}$. The procedure of finding unknown
frequencies $\omega_k$ and corresponding amplitudes $d_k=|\alpha_k|^2$ starts
from measured values of the time autocorrelation function
(\ref{autocorrelation}). In the autocorrelation function the
phases of the coefficients $\alpha_k$ cancel out and only the modulai $d_k$ enter 
the Eq. (\ref{autocorrelation}). Thus, in the definition of $P$ one can safely
use $\sqrt{d_k}$ instead of $\alpha_k$ and
\begin{equation}
P_{kn} \equiv \sqrt{d_k} e^{-i\omega_k n\delta t}.
\end{equation}

$P$ is a rectangular matrix, in general, and its condition number is expressed
in terms of singular values rather then eigenvalues, $\kappa(P)=p_1/p_K$, where
$p_1$ and $p_K$ are the largest and the smallest singular values of $P$,
respectively. 
Notice that the matrix $P$ has exactly $K$ singular values because there
are exactly $K$ linearly independent states $|k\rangle$ spanning the initial
state in Eq. (\ref{initstate}). The singular values $p_i$ are the square roots of
the singular values of $P^\dagger P=S$. That is $p_i=\sqrt{\lambda_i}$.
Finally,
\begin{equation}
\kappa(P) = \sqrt{\frac{\lambda_1}{\lambda_{min}}} \le \sqrt{\frac{\mbox{Tr}(S)}{\lambda_{min}}}.
\label{kappa}
\end{equation}
At that point the inequality (\ref{ieq1}) becomes useful because the upper
bound is expressed in terms of the trace and the smallest
positive eigenvalue of the known matrix $S$. The eigenvalue is discussed in the
following section.

\section{The smallest positive eigenvalue of $S$}
The problem of the error estimates of the extracted frequencies is completed in Eq.
(\ref{ieq1}) and Eq. (\ref{kappa}). To better understand the relation between
the possible error and the system parameters in Eq. (\ref{autocorrelation}), I will
express the eigenvalue $\lambda_{min}$ in terms of $K$, $N$, $T$, $\omega_k$ and $d_k$. 

It turns out that $\lambda_{min}$ can be analytically estimated in the regime
of short time span of the autocorrelation function. 
First helpful observation is that $\lambda_{min}$ is a very small positive number in the
limit where $\Delta\omega_{max} \delta t \ll 1$. The $S$ is Hermitian non-negative defined and it's
matrix elements $S_{mn}=c_{n-m}$ approach $\sum_{k=1}^Nd_k=\mbox{Tr}(S)/N$ if the time step
$\delta t$ is small. That means that the largest eigenvalue $\lambda_1$
approaches $\mbox{Tr}(S)$ and there is little room left for other eigenvalues
since the sum of all eigenvalues is $\mbox{Tr}(S)$. 
Second hint comes from the property that the nonzero eigenvalues of the $N\times N$ 
matrix $S=P^\dagger P$ are equal to nonzero eigenvalues of the $K\times K$ matrix 
$PP^\dagger$ since both are squares of singular values of the $K\times N$ matrix $P$. 
The last is a direct consequence of the singular value decomposition theorem \cite{Golub}.

Having established the smallness of $\lambda_{min}$ and the equivalence of
nonzero eigenvalues of $P^\dagger P$ and $PP^\dagger$, one sees that the
characteristic equation for $S$
\begin{equation}
a_N \lambda^N +...+ a_1\lambda + a_0=0
\end{equation}
gives the estimate $\lambda_{min}\approx -\frac{a0}{a1}$, where
$a_0=\det(PP^\dagger)$ and $a_1=-\sum_{k=1}^K\mbox{Minor}_{kk}(PP^\dagger)$.
The matrix $P$ has the following structure 
\begin{equation}
P=\mbox{diag}(\sqrt{d_1}, ..., \sqrt{d_K})\times V,
\end{equation}
where the $n$-th column of the $K\times N$ Vandermonde matrix $V$ is given by the vector 
$[e^{-i\omega_1 (n-1)\delta t}, ..., e^{-i\omega_K (n-1)\delta t}]$.
This leads to the determinant 
\begin{equation}
\det(PP^\dagger) = \det(VV^\dagger) \prod_{k=1}^K d_k.
\label{det}
\end{equation}
The determinant of the product $VV^\dagger$ of the Vandermonde matrices in the limit of
short time steps $\delta t$ is given by
\begin{eqnarray}
\det(VV^\dagger)&\approx &\left( \frac{N}{K}\right)^K \left[\prod_{j=1}^{K-1}
\left( \frac{N^2-j^2}{K^2-j^2}\right)^{K-j}\right] \times \nonumber\\
&\times & \delta t^{K(K-1)}\prod_{i,j>i=1}^K (\omega_j-\omega_i)^2.
\label{VandermondeDet}
\end{eqnarray}
The proof of this approximation is beyond the scope of this Letter. It is
easy, however, to numerically confirm correctness of Eq. (\ref{VandermondeDet}).

A minor $\mbox{Minor}_{kk}(PP^\dagger)$ is also given by Eq. (\ref{det}) where  $k$-th
amplitude is dropped and $k$-th frequency in Eq. (\ref{VandermondeDet}) is
dropped and also  $K$ is replaced with $K-1$ in both Eq. (\ref{det}) and Eq.
(\ref{VandermondeDet}).

Summarizing the calculations, the smallest eigenvalue of $S$ is expressed in terms of unknown
frequencies and amplitudes as follows
\begin{eqnarray}
\lambda_{min} &\approx& \frac{(N+K-1)!}{(N-K)!(2K-1)}\left[\frac{(K-1)!}{(2K-2)!}\right]^2\times
\nonumber\\
&\times& \frac{\delta t^{2(K-1)}}{\sum_{k=1}^K
\left[d_k\prod_{j=1,j\ne k}^K(\omega_k-\omega_j)^2\right]^{-1}}
\label{lambda}
\end{eqnarray}
for $K>2$, and
\begin{equation}
\lambda_{min} \approx \frac{N-1/N}{12}
\frac{(\omega_1-\omega_2)^2}{\frac{1}{d_1}+\frac{1}{d_2}} [N\delta t]^2
\label{lambdaK2}
\end{equation}
for $K=2$.

\section{Discussion of the results}
Despite the apparent complexity, the estimate of $\lambda_{min}$ has a simple
structure. To see this, it is convenient to first define the {\it effective frequency distance
$\Delta$}
\begin{equation}
\frac{1}{\Delta^{2(K-1)}}\equiv \frac{1}{K}\sum_{k=1}^K
\frac{d_1+...+d_K}{d_k\prod_{j=1,j\ne k}^K(\omega_k-\omega_j)^2}
\end{equation}
for $K>2$ and
\begin{equation}
\frac{1}{\Delta^2} \equiv
\frac{1}{2}\frac{1}{d_1d_2(\omega_1-\omega_2)^2},\qquad K=2.
\end{equation}
For $d_k=1$ the effective frequency distance 
$$\Delta\in \left[\min_{k,j}\{\omega_k-\omega_j\}, \max_{k,j}\{\omega_k-\omega_j\}\right]$$
with the tendency to approach the left end of the interval, meaning that
$\Delta$ characterizes the shortest frequency gaps.
The dependence on the amplitudes $d_k$ suggests that $\Delta$ is dominated also by the
weakest Fourier components of the autocorrelation function Eq.
(\ref{autocorrelation}).

The most important feature of $\lambda_{min}$ is that 
$\lambda_{min}\propto (N\delta t\Delta)^{2(K-1)}$. Using the total time span
$T=N\delta t$ one arrives at
\begin{equation}
\lambda_{min} \propto \left(T\Delta\right)^{2(K-1)}.
\end{equation}
Since only short measurements are considered $T\Delta\omega_{max}\ll 1$ it follows
that $T\Delta\ll 1$ and $\lambda_{min}$ decreases exponentially fast with the
dimensionality of the Hilbert space the state $|\psi\rangle$ lives in. That
in turn implies that the upper bound on the measured frequency 
inaccuracy in Eq. (\ref{ieq1}) is exponentially sensitive to the number of
Fourier components $K$ in Eq. (\ref{autocorrelation}). 
Thus, for very short measurement times $T\Delta\ll 1$, even a
moderate number of frequencies $K$ can blow the corresponding inaccuracies to
a very high level. And this is perhaps the intuition that is hidden behind
many time-energy uncertainty relations: ``One cannot measure frequencies
accurately in a very short time.'' 
This intuition does not reflect a fundamental principle. It merely states the practical difficulty
in analyzing systems involving many frequencies. 

Although there are time-energy uncertainty principles that are not
based on an intuition but on a rigorous derivation from equations of quantum mechanics
\cite{Mandelshtam}, their meaning is different from the one considered in this
work.

In fact one could perform an experiment testing the absence of the lower bound
on the energy measurement accuracy in a very short time, which is set by some
time-energy uncertainty relations. Such an experiment
would involve many copies of a two level quantum system prepared in 
the same state. Many copies are necessary to decrease the statistical noise
$\eta_{max}$ and the two energy levels would allow to keep $\lambda_{min}$ at
a reasonable value. At fixed time interval $T$ it is possible to adjust
$\lambda_{min}$, as long as $\lambda_{min}\ll 1$, by varying the number of
time steps $N$, see Eq. (\ref{lambdaK2}).

\section{Summary}
In this work the energy measurement on an evolving quantum system is
considered. The evolution is governed by a discrete Hamiltonian and the
initial state is spanned on a finite number $K$ of significant energy eigenstates. 
The main result is the following upper bound
\begin{equation}
|\tilde \omega_k - \omega_k|T \le
\frac{KN(N+1)\sqrt{\mbox{Tr(S)}}}{\lambda_{min}^{3/2}}\eta_{max}.
\label{ieq2}
\end{equation}
for inaccuracy $|\tilde \omega_k -\omega_k|$ of measured frequency $\tilde \omega_k$ with respect
to the actual frequency $\omega_k$. The total measurement time $T$ is short
as compared to the characteristic time scales of the system, $T\omega_{max} \ll 1$, and
is divided into $N$ time intervals by the sampling measurements.
If the noise $\eta_n$ is statistical then its maximal level $\eta_{max}$ can be reduced by
simultaneous use of many copies of the system. The systematic errors can, in
principle, be eliminated.
Despite the short measurement time $T$ the upper bound can be made arbitrarily low
by the reduction of $\eta_{max}$.

\section{Acknowledgments}
I would like to thank Jacek Dziarmaga, Krzysztof Sacha and George Zweig for
many discussions on the time-energy uncertainty relations. This work was
supported by LDRD program X1F3.

\end{document}